\documentclass[reprint,aps,prc,twocolumn,superscriptaddress]{revtex4-1}
\usepackage{CJK}
\usepackage{graphicx}% Include figure files
\usepackage{hyperref}
\usepackage{amssymb}
\usepackage{amsmath}
\usepackage[normalem]{ulem}
\usepackage{url}
\usepackage{color}
\begin{document}
\begin{CJK*} {UTF8}{} %{GB} {gbsn}
%\preprint{APS/123-QED}

\title{How Do Constraints of Nuclear Symmetry Energy Reconcile with Different Models?}

\author{Yingxun Zhang}
\email{zhyx@ciae.ac.cn}
\affiliation{China Institute of Atomic Energy, Beijing 102413, China}
\affiliation{Guangxi Key Laboratory of Nuclear Physics and Technology, Guangxi Normal University, Guilin, 541004, China}

\author{Yangyang Liu}
\email{liuyangyang@ciae.ac.cn}
\affiliation{China Institute of Atomic Energy, Beijing 102413, China}

\author{Yongjia Wang}
\email{wangyongjia@zjhu.edu.cn}
\affiliation{School of Science, Huzhou University, Huzhou 313000, China}

\author{Qingfeng Li}
\email{liqf@zjhu.edu.cn}
\affiliation{School of Science, Huzhou University, Huzhou 313000, China}
\affiliation{Institute of Modern Physics, Chinese Academy of Sciences, Lanzhou 730000, China}

%\author{Ying Cui}
%\affiliation{China Institute of Atomic Energy, Beijing 102413, China}
\author{Zhuxia Li}
\affiliation{China Institute of Atomic Energy, Beijing 102413, China}
%\author{Yongjing Chen}
%\affiliation{China Institute of Atomic Energy, Beijing 102413, China}

\date{\today}

\begin{abstract}
%Based on the ultra-relativistic quantum molecular dynamics (UrQMD) model, the isospin sensitive flow and pion observables for $Au+Au$ collisions at 0.4A GeV are investigated. %Our analysis showed that the flow characteristic density is around 1.2$\rho_0$ and pion characteristic density is around 1.5$\rho_0$. 
By simultaneously describing the data of isospin sensitive nucleonic flow and pion observables, such as $v_2^n/v_2^{ch}$ and $\pi^-/\pi^+$, with ultra-relativistic quantum molecular dynamics (UrQMD) model, we got the symmetry energy at flow and pion characteristic densities which are $S(1.2\rho_0)=34\pm 4$ MeV and $S(1.5\rho_0)=36\pm 8$ MeV. Within the uncertainties, the constraints of symmetry energy at characteristic densities are consistent with the previous constraints by using other transport models. The consistency suggests that the reliable constraints on symmetry energy should be presented at the characteristic density of isospin sensitive observables. By using the constraints of symmetry energy at two different characteristic densities, the extrapolated value of $L$ is provided. Within $2\sigma$ uncertainty, the extrapolated value of $L$ is in $5-70$ MeV which is consistent with the recent combination analysis from PREX-II and astrophyiscs data. Further, the calculations with the constrained parameter sets can describe the data of charged pion multiplicities from S$\pi$RIT collaboration. 

%By performing the calculations with the constrained parameter sets, the charged pion multiplicities and ratios obtained with S$\pi$RIT can also be described.

%Different than the model dependent results on the extrapolated constraints of the slope of symmetry energy at normal density $L$, 

%produced from the $^{197}Au+^{197}Au$ collision at the beam energy $E_{beam}=0.4A$ GeV
\end{abstract}

%\keywords{}

\maketitle
\end{CJK*}

%\section{Introduction}
%\label{introduction}
%\emph{Introduction}:
Knowledge of the symmetry energy is crucial for understanding the isospin asymmetric objects, such as the structure of neutron-rich nuclei, the mechanism of neutron-rich heavy ion collisions, and the properties of neutron stars\cite{BALi08,CJHorowitz2014JPG}. However, theoretical predictions on the symmetry energy away from the normal density show larger uncertainty, and it leads that the constraint of symmetry energy becomes one of the important goals in nuclear physics\cite{Carlson2017,NuPECC17}.

For probing the symmetry energy at suprasaturation density, the isospin-sensitive observables in heavy ion collisions (HICs), such as the ratio of elliptic flow of neutron to charged particle, hydrogen
isotopes or proton ($v^n_2/v^{ch}_2$, $v^n_2/v^{H}_2$ or $v^n_2/v^{p}_2$)\cite{Russotto2011PLB,Cozma2013PRC,WangYJ2014PRC,Russotto2016PRC,Cozma2018EPJ} and the yield ratios of charged pions (i.e., $M(\pi^-)/M(\pi^+)$ or named as $\pi^-/\pi^+$)\cite{BALi2002PRL,BALi2002NPA,ZGXiao2009PRL,ZQFeng2010PLB,WJXie2013PLB,JH2014PRC,Song15,Cozma2016PLB,YYLiu2021PRC,GCYong2021PRC,SpiRIT2021PRL}, were frequently used. By comparing the transverse-momentum-dependent or integrated FOPI/LAND
elliptic flow data of nucleons and hydrogen isotopes with UrQMD\cite{Russotto2011PLB,WangYJ2014PRC} and T\"uQMD models\cite{Cozma2013PRC}, a moderately soft to linear symmetry energy is obtained\cite{Russotto2016PRC}. Usually, the constraints of symmetry energy were presented by the symmetry energy coefficient $S_0$ and the slope of symmetry energy $L$ at normal density. The lower limit of $L$ obtained with the flow ratio data is $60$ MeV\cite{Wang2020Fronp}, which overlaps with the upper limits of the constraints from the nuclear structure and isospin diffusion, i.e., $L\approx 60\pm 20$ MeV\cite{BALi13,Oertel17,YXZhang2020PRC}. The deduced $L$ from FOPI $\pi^-/\pi^+$ data show strong model dependence\cite{ZGXiao2009PRL,ZQFeng2010PLB,WJXie2013PLB,JH2014PRC,Song15,Cozma2016PLB,YYLiu2021PRC,GJhang20}, which ranges from 12 MeV to 144 MeV.% due to the different treatments in the mean field part and nucleon-nucleon collision part.  provide constraint on symmetry energy at suprasaturation density by comparing the data with dcQMD (previously named T\"uQMD)\cite{Cozma2021epja}, and the extrapolated slope of symmetry energy is $42 < L < 117$ MeV. 

%It may be caused by the different treatments on the $\Delta$ potential, threshold effects, pion potential, Pauli blocking, in-medium cross sections and so on, and also by the different numerical technical for solving the transport equations.

% weak sensitivity of $\pi^-/\pi^+$ to the slope of symmetry energy\cite{JH2014PRC,YYLiu2021PRC}.
%Recently, the PREX-II experiment reported a new measurement of the neutron-skin of $^{208}$Pb, and led the Ref.\cite{Reed2021} conclude that $L=106\pm37$ MeV, This value is larger than the previous results and present a challenge to our understanding on the symmetry energy.  the Ref.\cite{Tong2021} conclude that $L=85.5\pm 22.2$ MeV.
%\cite{Akira2019PRC,YXZhang2018PRC,Maria2021PRC,JunXu2016PRC}

To reduce the model dependence on the constraints of symmetry energy, especially at suprasaturation densities, the transport model evaluation project was performed among the transport model community since 2009. Up to now, the transport model evaluation project has made important progress on the reduction of model dependence of transport model by benchmarking the treatment of particle-particle collision\cite{YXZhang2018PRC,Akira2019PRC} and nucleonic mean field potential\cite{Maria2021PRC} in both Boltzmann-Uehling-Uhlenbeck (BUU) type and Quantum molecular dynamics (QMD) type models. However, the recent works on S$\pi$RIT pion data \cite{SpiRIT2021PRL} for Sn+Sn collision system at 0.27A GeV have also shown the model dependence after comparing the pion production with seven updated transport models. It hints the model dependence of the inferred $L$ by using flow or pion observables may come from the extrapolation of symmetry energy from the probed density to normal density. %, and thus on the constrained values of $L$.%  The recent published S$\pi$RIT pion data \cite{SpiRIT2021PRL} for Sn+Sn at 270A MeV provided a new opportunity for constraining the symmetry energy at suprasaturation density and benchmark the transport model. Seven transport models have been used to compare with the data, but the model dependence was still there\cite{GJhang20}.  

To verify this issue, one has to know which density is probed by flow and pion observables. The named characteristic density was proposed in Refs.~\cite{YYLiu2021PRC,Lynch2022PLB,Fevre2016NPA,YXZhang2020PRC}, which is used to quantitatively describe the density probed by the flow or pion observables. After knowing the characteristic density, the consistency or inconsistency of symmetry energy constraints can be further understood. %To do that, the characteristic density of flow and pion observables should be investigated first. There were some efforts to calculate the characteristic density for flow observables\cite{Fevre2016NPA} and pion observable\textcolor{red}{\cite{YYLiu2021PRC,HLLiu2015PRC}}. 
In this work, we simultaneously investigate the flow and pion observables within the framework of UrQMD model. By comparing the calculation to ASY-EOS and FOPI data, the data favored parameter sets are obtained and a remarkable consistency is found at both flow and pion characteristic densities among different transport models. %The consistency suggests that the reliable constraints on symmetry energy should be presented at the characteristic density of isospin sensitive observables rather than at normal denisty. In addition, to extrapolate the value of $L$ at normal density, a better way is to use the constraints of symmetry energy at least two characteristic densities. %The characteristic density of flow observables is discussed first.

The version of UrQMD model we used is the same as that in Ref.\cite{YYLiu2021PRC} but the momentum dependent interaction and the cross sections of $NN\to N\Delta$ near the threshold energy are refined, more details are in Ref.~\cite{YYLiu2022arXiv}. %, in which the cross sections of $N\Delta \to NN$ are replaced with a more delicate form by considering the $\Delta$-mass dependence of the M-matrix in the calculation of $N\Delta \to NN$\cite{YCui2019CPC}. 
%This version has been successfully used to describe the FOPI experimental data of multiplicity and ratio of charged pion\cite{YYLiu2021PRC}. 
%In the model, each nucleon is represented by a Gaussian wave packet. The centriods of a nucleon in the coordinate and momentum space, i.e., $\textbf{r}_i$ and $\textbf{p}_i$, are propagated under the mean-field potential and the nucleon-nucleon collision.
%For mean field part, $\textbf{r}_i$ and $\textbf{p}_i$ evolve according to the Hamiltonian equation,
%\begin{equation}\label{eqHamilton}
% \dot{\textbf{r}}_i=\frac{\partial H }{\partial \textbf{p}_i},\dot{\textbf{p}}_i=-\frac{\partial H }{\partial \textbf{r}_i}.
%\end{equation}
%The Hamiltonian $H$ consists of the kinetic energy $T$ and the effective interaction potential energy $U$.
The nucleonic potential energy $U$ is calculated from the potential energy density, i.e., $U=\int u d^3 r$. The $u$ reads as
 %with $T=\sum_i(E_i-m_i)=\sum_i(\sqrt{m_i^2+p_i^2}-m_i)$ and $U=U_\rho+U_{md}+U_{coul}$. $U(\rho)=\int u(\rho) d\textbf{r}$.
\begin{eqnarray}
u & =& \frac{\alpha}{2}\frac{\rho^2}{\rho_0}+\frac{\beta}{\eta+1}\frac{\rho^{\eta+1}}{\rho_0^\eta}\\\nonumber
&& +\frac{g_{sur}}{2\rho_0}(\nabla \rho)^2+\frac{g_{sur,iso}}{\rho_0}[\nabla(\rho_n-\rho_p)]^2\\\nonumber
&&+u_{md}+u_{sym}.
\end{eqnarray}
The parameters $\alpha$, $\beta$, and $\eta$ are related to the two, three-body interaction term. The third and fourth terms are isospin independent and isospin dependent surface term, respectively. The $u_{md}$ is from the momentum dependent interaction (MDI) term, which is calculated according to the following relationship,
\begin{equation}
\begin{aligned}
u_{m d}=& \sum_{N_1,N_2=n,p}\int d^{3} p_{1} d^{3} p_{2} f_{N_1}\left(r,\vec{p}_{1}\right) f_{N_2}\left(r,\vec{p}_{2}\right)\\
& v_{md}(\Delta p_{12}).
\end{aligned}
\end{equation}
$f_{N_i}(r,\vec{p})$ is the phase space density of particle $N_i$. The form of MDI, i.e., $v_{md}(\Delta p_{12})$, is assumed as,
\begin{equation}\label{Vmd-p}
	v_{md}(\Delta p_{12})= t_4\ln^2(1+t_5\Delta p_{12}^2)+c,
\end{equation}
where $\Delta p_{12}=|\textbf{p}_1-\textbf{p}_2|$, and the parameters $t_4$, $t_5$  and c are obtained by reproducing the real part of the optical potential of Hama's data by using Eq.(\ref{Vmd-p}) and
\begin{equation}\label{Vmdvmd}
   V_{md}(p_1)=\int_{p_2<p_F}v_{md}(p_1-p_2)d^3p_2/\int_{p_2<p_F}d^3p_2,
\end{equation}
within the kinetic energy $E_{kin}\approx$ 750 MeV. The method we used is as same as in Ref.\cite{Hartnack1994prc}. 
With the consideration of MDI, the parameter $\alpha$, $\beta$, and $\eta$ are readjusted to keep the incompressibility of symmetric nuclear matter $K_0=231$ MeV. The values of parameters and corresponding effective mass are listed in Table~\ref{tab:table1}.

\begin{table}[htbp]%The best place to locate the table environment is directly after its first reference in text
\caption{\label{tab:table1}%
Parameters used in this work.  $t_4$, $c$, $\alpha$, $\beta$ and $K_0$ are in MeV. $t_5$ is in MeV$^{-2}$. $\eta$ and $m^*/m$ are dimensionless.}
%\begin{ruledtabular}
%\centering
\begin{tabular}{lcccccccc}
\hline
\hline
$Para.$ & $t_4$ & $t_5$ & $c$ &$\alpha$ & $\beta$ & $\eta$  & $K_0$ & $m^*/m$\\
\hline
%\multicolumn{1}{c}{\textrm{Decimal}}&
%\colrule
%$v_{md}^{Arnold}$ & 1.57 & 5$\times$10$^{-4}$ & -54  &-221 & 153 & 1.31   & 231 & 0.77 \\
%\hline
$v_{md}^{Hama}$ & 3.058 & 5$\times$10$^{-4}$ & -86  &-335 & 253 & 1.16  & 231 & 0.635 \\
\hline
\hline
\end{tabular}
%\end{ruledtabular}
\end{table}

For the potential energy density of symmetry energy part, i.e., $u_{sym}$, we take two forms in the calculations. One is the Skyrme-type polynomial form (form (a) in Eq.~(\ref{srho-lyy})) and another is the density power law form (form (b) in Eq.~(\ref{srho-lyy})). It reads,
\begin{eqnarray}
\label{srho-lyy}
 u_{sym}&=&S^{pot}_{sym}(\rho)\rho\delta^2\\\nonumber
 &=&\left\{
 \begin{array}{ll}
    ( A(\frac{\rho}{\rho_0})+B(\frac{\rho}{\rho_0})^{\gamma_s}+C(\frac{\rho}{\rho_0})^{5/3} )\rho\delta^{2}, & \mathbf{(a)}\\
    \frac{C_{s}}{2}(\frac{\rho}{\rho_{0}})^{\gamma_i}\rho\delta^2. & \mathbf{(b)}
  \end{array}
\right.
\end{eqnarray}
%In our calculations, we $S_0$ varies from 30 to 34 MeV, and $L$ varies from 5 to 144 MeV. The simple power law form of symmetry energy is used for $L>35$ MeV, and the parameters $C_s$ and $\gamma_s$ are determined by the $S_0=S(\rho_0)$ and  $L=3\rho_0\frac{\partial S(\rho)}{\partial \rho}|_{\rho=\rho_0} $. For $L<35$ MeV, this form can not give reasonable symmetry energy at subnormal density, and thus we use the Skyrme polynomial form of $S(\rho)$ in the calculations for $L$ = 5, 20, and 35 MeV. The parameters $A_{sym}$, $B_{sym}$, $C_{sym}$ and $\gamma_s$ are obtained based the five nuclear matter parameters $K_0$, $S_0$, $L$, $m^*_s/m$, and $f_I$, at given values of $\rho_0$, $E_0$ and $g_{sur}$\cite{YXZhang2020PRC}. The $L < 5$ MeV sets are not adopted, because the corresponding symmetry energy becomes negative at the densities above $2.7\rho_0$ and the EOS will not be favored by the properties of the neutron star. $S_0$ is taken as $30$, $32.5$ and $34$ MeV. If there is no specific explanation, $S_0$ is taken as 32.5 MeV in the following discussions.
In the following calculations, $S_0$ varies from 30 to 34 MeV and $L$ varies from 5 to 144 MeV. The corresponding parameters in Eq.(\ref{srho-lyy}) are determined with the relationship in Ref.\cite{YXZhang2020PRC,Zhang2020FOP}.

The calculations of Au+Au at 0.4A GeV and Sn+Sn at 0.27 A GeV are performed with 200,000 events. Since the collision centrality is determined by the $Z_{bound}$ or $Z_{rat}$ in the Ref.~\cite{Russotto2016PRC} and the detected charged particle multiplicity~\cite{GJhang20} or the ratio of total transverse to longitudinal kinetic energies in the center-of-mass (c.m.) system in Ref.~\cite{FOPI2010}, the corresponding impact parameter $b$ will be in a wide range and the weight of $b$ is a Gaussian shape rather than a triangular shape. This impact smearing effect~\cite{Russotto2016PRC,Frankland2021PRC,Lili2019PRC,Lili2022arxiv} is also considered in calculations. In Table.\ref{tab:observable}, we list the experimental sixteen observables we investigated. 
%which are as same as in experiments\cite{Russotto2016PRC}}.
\begin{table}[htbp]%The best place to locate the table environment is directly after its first reference in text
\caption{\label{tab:observable}%
Sixteen experimental observables and the corresponding averaged impact parameters (in fm) analyzed in this work.}
%\begin{ruledtabular}
%\centering
\begin{tabular}{lcccc}
\hline
\hline
$\text{Obsevable}$ &  $<b>$ & system & Exp.\\
\hline
%\multicolumn{1}{c}{\textrm{Decimal}}&
%\colrule
$v^n_1(p_t/A)$ & 5.69 & Au+Au & ASY-EOS~\cite{Russotto2016PRC} \\
%\hline
$v^{ch}_1(p_t/A)$ &  5.69 & Au+Au & ASY-EOS~\cite{Russotto2016PRC}\\
$v^{n}_2(p_t/A)$ &  5.69  & Au+Au & ASY-EOS~\cite{Russotto2016PRC}\\
$v^{ch}_2(p_t/A)$ &  5.69  & Au+Au & ASY-EOS~\cite{Russotto2016PRC}\\
$v_2^n/v_2^{ch}(p_t/A)$ &  5.69 & Au+Au & ASY-EOS~\cite{Russotto2016PRC}\\
$M(\pi)$ &  $<$2\footnote{We did not put the average b value here since experimental paper only provides $b/b_{max}<0.15$, which is obtained by estimating the impact parameter b from the measured differential cross sections for the
ERAT under a geometrical sharp-cut approximation. \label{1}}  & Au+Au & FOPI~\cite{FOPI2010}\\
$\pi^-/\pi^+$ &  $<$2\textsuperscript{\ref {1}}.  & Au+Au & FOPI~\cite{FOPI2010}\\
$Y(\pi^-)$ &  3  & $^{108}$Sn+$^{112}$Sn & S$\pi$RIT~\cite{GJhang20}\\
$Y(\pi^-)$ &  3  & $^{112}$Sn+$^{124}$Sn & S$\pi$RIT~\cite{GJhang20}\\
$Y(\pi^-)$ &  3  & $^{132}$Sn+$^{124}$Sn & S$\pi$RIT~\cite{GJhang20}\\
$Y(\pi^+)$ &  3  & $^{108}$Sn+$^{112}$Sn & S$\pi$RIT~\cite{GJhang20}\\
$Y(\pi^+)$ &  3  & $^{112}$Sn+$^{124}$Sn & S$\pi$RIT~\cite{GJhang20}\\
$Y(\pi^+)$ &  3  & $^{132}$Sn+$^{124}$Sn & S$\pi$RIT~\cite{GJhang20}\\
$\pi^-/\pi^+$ &  3  & $^{108}$Sn+$^{112}$Sn & S$\pi$RIT~\cite{GJhang20}\\
$\pi^-/\pi^+$ &  3  & $^{112}$Sn+$^{124}$Sn & S$\pi$RIT~\cite{GJhang20}\\
$\pi^-/\pi^+$ &  3  & $^{132}$Sn+$^{124}$Sn & S$\pi$RIT~\cite{GJhang20}\\
\hline
\hline
\end{tabular}
%\end{ruledtabular}
\end{table}

\begin{figure}[htbp]
\centering
\includegraphics[angle=0,scale=0.35]{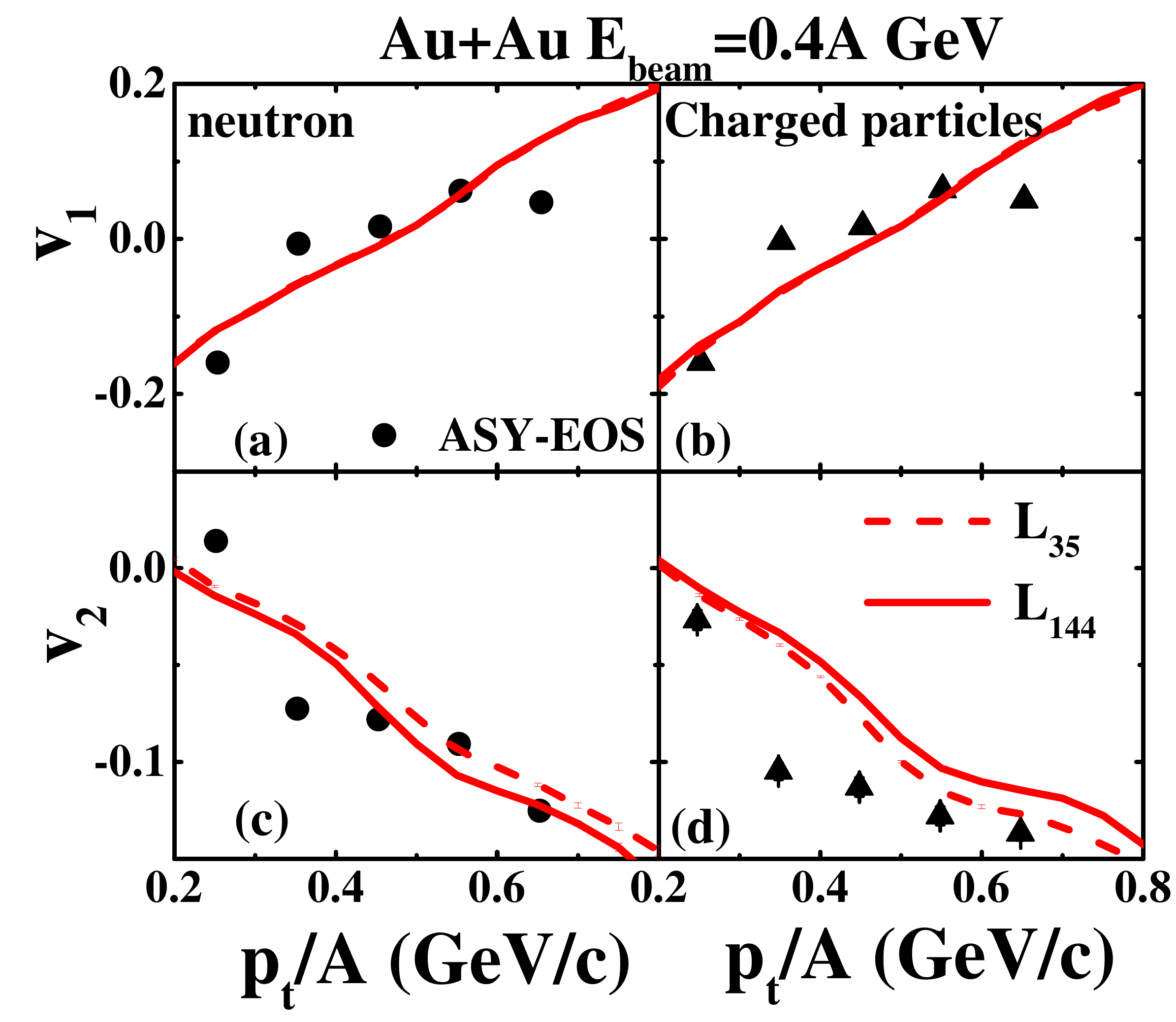}
\setlength{\abovecaptionskip}{0pt}
\vspace{2em}
\caption{The directed flow $v_1$ (upper panels) and elliptic flow $v_2$ (lower panels) of free neutrons (left panels) and charged particles (right panels). The dash and solid lines represent the results with $L=35$ MeV and $L=144$ MeV, respectively. The ASY-EOS data of collective flow for neutron and charged particles are shown as circle and triangle symbols\cite{Russotto2016PRC}.}
\setlength{\belowcaptionskip}{0pt}
\label{collective-flow}
\end{figure}

Fig.\ref{collective-flow} illustrates the UrQMD model calculations and the measured data for the directed ($v_1$) and the elliptic flow ($v_2$) of neutrons and charged particles as a function of the transverse momentum per nucleon $p_t/A$. The intervals of the polar angle and rapidity are chosen to be the same as in the ASY-EOS analysis, i.e., $37^ {\circ}<\theta_{lab}< 53^ {\circ}$ and $-0.5<y_0<0.5$. As an example, we present the results obtained with $L=35$ MeV (dashed lines) and $L=144$ MeV (solid lines) at $S_0=30$ MeV. The symbols are the ASY-EOS data from Ref.\cite{Russotto2016PRC}. As shown in panels (a) and (b), the calculations with different $L$ fall in the data region. But the results calculated with $L=35$ MeV and $L=144$ MeV sets are overlapping each other, which indicates $v_1$ in the plotted intervals has no sensitivity to the nuclear symmetry energy, due to the spectator matter blocking effect. 

%More details, the $v_1^n(p_t/A)$ and $v_1^{ch}(p_t/A)$ increase from negative values to positive values with the increasing of $p_t/A$, and the sign of $v_1$ changes around $p_t/A\approx 0.5$ GeV/$c$. 

%For the directed flow, 

\begin{figure}[htbp]
	\centering
	\includegraphics[angle=0,scale=0.48]{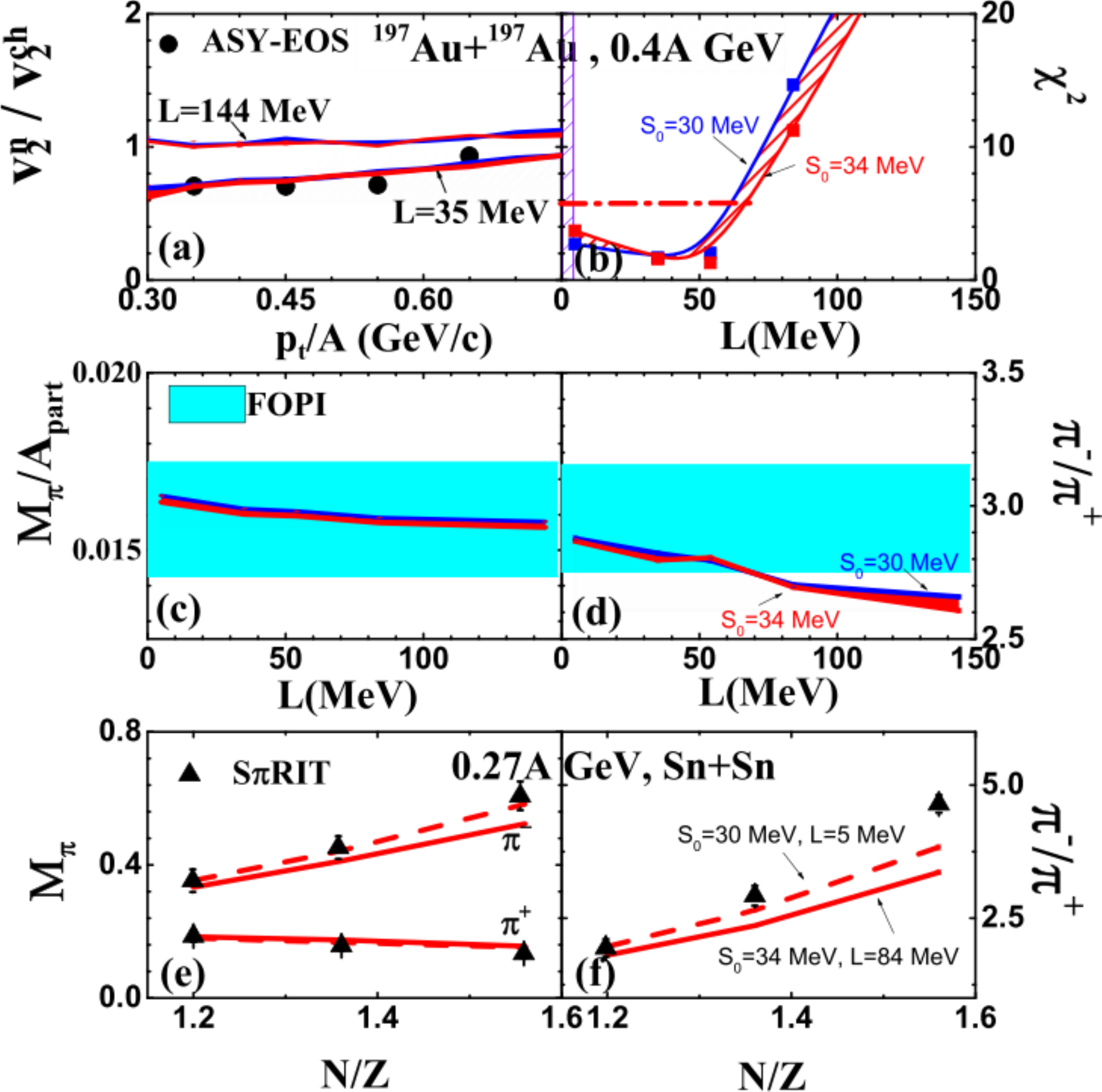}
	\setlength{\abovecaptionskip}{0pt}
	\vspace{2em}
	\caption{Panel (a) $v_2^n/v_2^{ch}$ as a function of $p_t$/A for $L=35$ MeV and $L=144$ MeV with $S_0$ in the range from 30 MeV to 34 MeV; (b) $\chi^2$ as a function of $L$ for $S_0$=30 (blue lines) and 34 MeV (red lines). Panel (c) and (d) are $M(\pi)/A_{part.}$ and $\pi^-/\pi^+$ ratio as a function of $L$ with $S_0=30$ and $34$ MeV, respectively. Panel (e) and (f) are the multiplicity of charged pion and its ratio as a function of $N/Z$ of the system. The black symbols represent the ASY-EOS experimental data\cite{Russotto2016PRC},  the cyan-shaded region is the FOPI data and the blue symbols are the S$\pi$RIT data.} 
	\setlength{\belowcaptionskip}{0pt}
	\label{flow-pion}
\end{figure}

For the elliptic flow, the effects of symmetry energy on $v_2$ of both free neutrons and charged particles can be clearly observed as in Fig.\ref{collective-flow} (c) and (d). Both the $v_2^n$ and $v_2^{ch}$ have negative values and decrease with $p_t/A$ increasing, which means a preference for particle emission out of the reaction plane, towards $90^\circ$ and $270^\circ$. As shown in Fig.\ref{collective-flow} (c), the values of $v_2^n$ obtained with stiff symmetry energy case are lower than that with soft symmetry energy case. The reason is that the stiff symmetry energy provides a stronger repulsive force on neutrons at suprasaturation density than that for soft symmetry energy cases. For charged particles, as shown in panel (d), $v_2^{ch}$ obtained with the stiff symmetry energy case are higher than that with the soft symmetry energy case. This is because the emitted charged particles are mainly composed of free protons, which feel stronger attractive interaction for the stiff symmetry energy case than that for the soft symmetry energy case at suprasaturation density. Consequently, $v_2^{ch}$ obtained with the stiff symmetry energy case is higher than that with the soft symmetry energy case, which has also been discussed in Refs.~\cite{Russotto2011PLB, Cozma2013PRC, WangYJ2014PRC,YJWang2020PLB}. However, $v_2^n$ or $v_2^{ch}$ cannot be used individually to constrain the symmetry energy, because both $v_2^n$ or $v_2^{ch}$ not only depend on the symmetry energy but also on the MDI and incompressibility. For example, the calculations with different incompressibility can lead to different results on the elliptic flow\cite{Wang2018PLB}.
To isolate the contributions from the isocalar potential, $v_2^n/v_2^{ch}$ ratio was proposed to probe symmetry energy and several analysis have been performed by using the UrQMD model or T\"uQMD model\cite{Russotto2016PRC,Cozma2016PLB}. In Fig.\ref{flow-pion} (a), we present the calculated results for $v_2^n/v_2^{ch}$ as a function of $p_t$/A obtained with $L=35$ MeV and $144$ MeV sets at $S_0=30$, $32.5$ and $34$ MeV. The symbols are the ASY-EOS data points. %The cuts of rapidity $-0.5<y_0<0.5$ and angle $37^{\circ}  < \theta_{lab.}< 53^{\circ} $ are also as the same as in experiment data\cite{Russotto2016PRC}.
The calculations show that $v_2^n/v_2^{ch}$ is sensitive to $L$, especially at the low $p_t$ region in which the mean-field play more important role. The values of $v_2^n/v_2^{ch}$ obtained with $L=144$ MeV sets are larger than that with $L=35$ MeV sets. This behavior can be understood from Fig.\ref{collective-flow} (c) and (d). By comparing the calculations of $v_2^n/v_2^{ch}$ to ASY-EOS experimental data and doing a $\chi^2$ analysis, one draw the conclusion on the slope of symmetry energy is 5-67 MeV for $S_0=32.5$ MeV case. After considering the uncertainties of $S_0$, the constraints on the $L$ extend to 5-70 MeV which are presented in panel (b). %The $L < 5$ MeV sets are not adopted, because the corresponding symmetry energy becomes negative for densities above 2.7$\rho_0$ and the EOS will not favor the neutron star.

Fig.\ref{flow-pion} (c) and (d) show the calculated $M_{\pi}$ and $\pi^-/\pi^+$ for Au+Au as a function of $L$ (shaded region with different color boundary) and the FOPI data (cyan shaded region). The width of the red-shaded region represents the results obtained with $S_0$ from 30 to 34 MeV. $M(\pi)$ calculated with different $S_0$ and $L$ falls into the data region, and suggest that $M(\pi)$ can not be used to distinguish $L$. The $\pi^-/\pi^+$ ratios obtained with UrQMD model are sensitive to $L$, and it decreases with the increase of $L$. %However, the symmetry energy effect is not so strong enough to distinguish the different $L$ with an high accuracy. %This is consistent with the results obtained with pBUU model\cite{JH2014PRC}. 
%One explanation is that the pions experience averaged 4.5 times $N-\Delta-\pi$ loops before freezing out\cite{YYLiu2021PRC}, and the subsequent interactions on the pion production tend to erase the effect of symmetry energy from the initial $\Delta$ decay occurring at high density. 
Comparing the calculated results of $\pi^-/\pi^+$ to the FOPI experimental data, the parameter sets with $L$=5-70 MeV are favored.

To test the validation of obtained parameter sets, we also perform the calculations of Sn+Sn and compare the pion production results to S$\pi$RIT data (black symbols). Fig.\ref{flow-pion} (e) and (f)  show the charged pion multiplicities and its ratio as a function of N/Z of system. Two extreme values of $L$ are used in the validation. One is for $S_0=30$ MeV and $L=5$ MeV and another is $S_0=34$ MeV and $L=84$ MeV. The testing calculations demonstrate that the calculation with $L=5$ MeV set can describe the $M(\pi^-)$ and $M(\pi^+)$ for three reaction system, $^{108}$Sn+$^{112}$Sn, $^{112}$Sn+$^{124}$Sn and $^{132}$Sn+$^{124}$Sn, within the experimental uncertainties. For $\pi^-/\pi^+$ ratios, the calculations can reproduce the data for the $^{108}$Sn+$^{112}$Sn and $^{112}$Sn+$^{124}$Sn system, but underestimate the data for very neutron rich system $^{132}$Sn+$^{124}$Sn. The discrepancy is related to the subthreshold pion production mechanism, where the  threshold effects\cite{Ferini2005NPA,TSong2015PRC,ZhenZhang2017PRC} and isospin-dependent medium correction on $NN\to N\Delta$ cross section\cite{QFLi2017PLB,YCui2018PRC} become important.
Although the data favored parameter sets are presented with $L$ value at normal density, one should keep in mind that the isospin observables $v_2^{n}/v_2^{ch}$ and $\pi^-/\pi^+$ probe the symmetry energy information at suprasaturation density region rather than at normal density.
The interesting point is what density is probed by flow and pion observables? Here, we use the named characteristic density to quantitatively describe them. For pion observable, the characteristic density is obtained by averaging the compressed density with pion production rate in spatio-temporal domain~\cite{YYLiu2021PRC}, and the calculations show that the characteristic density of pion observable is around 1.5 times normal density.
%Constraining the symmetry energy at the sensitive observable characteristic density is also important for having more precise knowledge of the symmetry energy\cite{Lynch2021}. 

For collective flow, the idea of calculating characteristic density is as same as pion characteristic density\cite{YYLiu2021PRC,Fevre2016NPA}, but the weight is replaced by the momentum change of nucleons which reflects strength of driven force for the collective motion of emitted particles. 
%The definition of flow characteristic density read as,
%\begin{equation}\label{rho-ch-px}
%\left\langle\rho_{\mathrm{c}, \text { flow }}\right\rangle_{\left|\Delta p_\alpha \right|}=\frac{\int_{t_0}^{t_1} \Sigma_i\left|\Delta p_\alpha ^i(t)/\Delta t\right| \rho_c(t) d t}{\int_{t_0}^{t_1} \Sigma_i\left|\Delta p_\alpha^i(t)/\Delta t\right| d t}.
%\end{equation}
%and
%\begin{equation}\label{rho-ch-pt}
%\left\langle\rho_{\mathrm{c}, \text { flow }}\right\rangle_{\left|\Delta p_t\right|}=\frac{\int_{t_0}^{t_1} \Sigma_i\left|\Delta p_t^i(t)/\Delta t\right| \rho_c(t) d t}{\int_{t_0}^{t_1} \Sigma_i\left|\Delta p_t^i(t)/\Delta t\right| d t}.
%\end{equation}
%The summation over $i$ runs over the emitted nucleons and charged particles, and $|\Delta p_\alpha(t)/\Delta t|=|(p_\alpha(t)-p_\alpha(t-\Delta t))/\Delta t|$. $\alpha$ is the components of momentum of emitted particles. 
With this weight definition, the more momentum change during the time evolution is, the more weight on the collective motion of nucleon is. Finally, the characteristic density for the collective flow are obtained, and it is $1.2\pm 0.6\rho_0$. It is consistent with the characteristic density obtained in the Ref.\cite{Fevre2016NPA}, but is smaller than the characteristic density obtained with pion observable.

%Figure.\ref{rho-dp-t} (a) shows the time evolution of the averaged central density $\rho_c(t)$ which is obtained in the spherical region with a radius of 3.35fm and centered at c.m. of the system. For semi-peripheral collision of Au+Au, the averaged central density beyond normal density from 8 fm/$c$ to 28 fm/$c$ and reaches maximum values of 1.8$\rho_0$ at 16 fm/$c$ with the interactions we adopted. Panel (b) shows momentum changes during the time interval as a function of time which is used as a weight in Eq.(\ref{rho-ch-px}) for calculating the characteristic density. Two kinds of momentum changes of emitted particles are used. One is the change of momentum in $x$-direction during the time interval, i.e., $|\Delta p_x(t)/\Delta t|$, and another is the changes of transverse momentum during the time interval, i.e., $|\Delta p_t(t)/\Delta t|$. They are used to understand the origins of $v_1$ and $v_2$. It clearly illustrates that the emitted particles are largely accelerated from 8 fm/$c$ to 28 fm/$c$ during the compressed and expansion phase.

%\begin{figure}[htbp]
%\centering
%\includegraphics[angle=0,scale=0.32]{Fig1-rho-dp-t.pdf}
%\setlength{\abovecaptionskip}{0pt}
%\vspace{2em}
%\caption{(a) Time evolution of the averaged density in the center of
%reaction system, (b) time evolution of momentum changes in $x$ direction $|\Delta p_x/\Delta t|$ and transverse direction $|\Delta p_t/\Delta t|$.}
%\setlength{\belowcaptionskip}{0pt}
%\label{rho-dp-t}
%\end{figure}

\begin{figure*}[htbp]
\centering
\includegraphics[angle=0,scale=0.6]{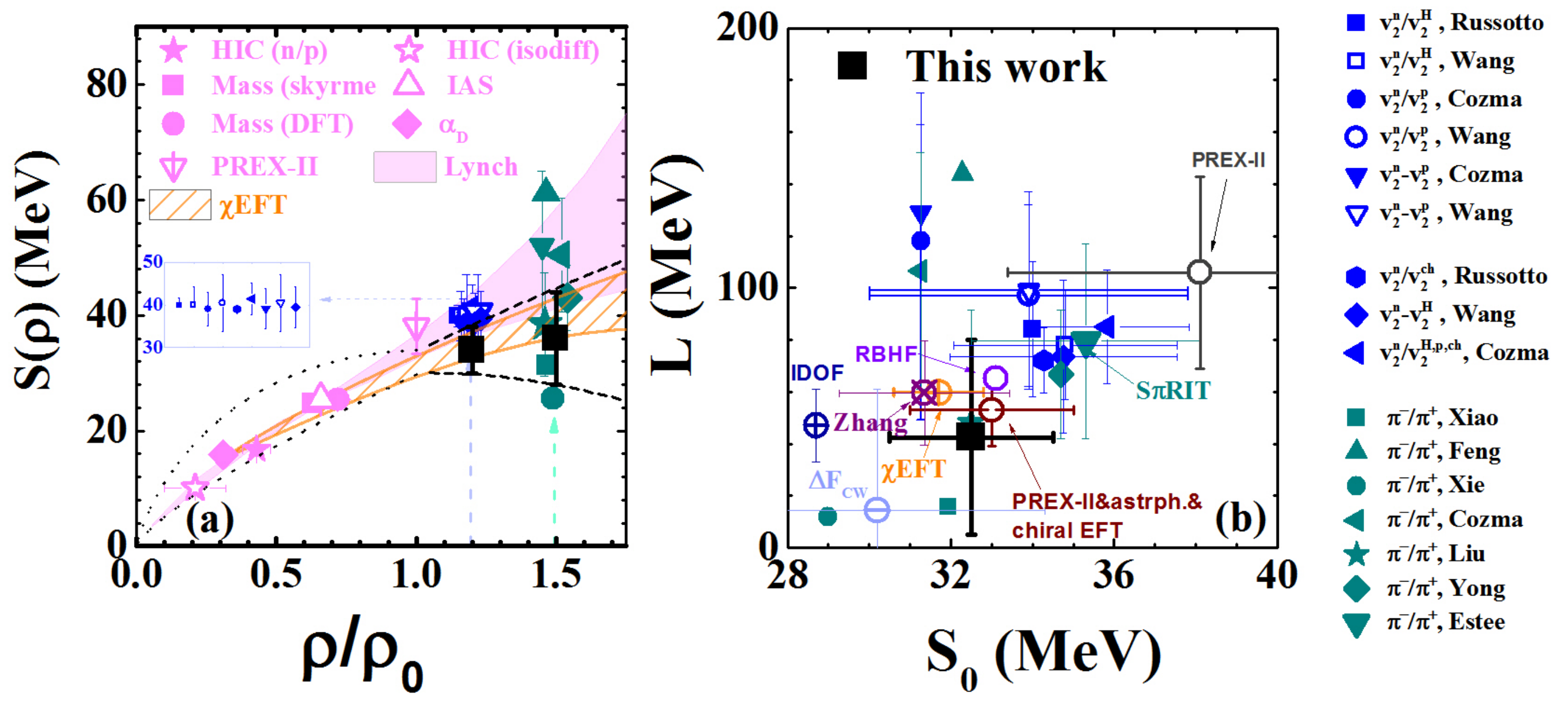}
\setlength{\abovecaptionskip}{0pt}
\vspace{2em}
\caption{Panel(a): The constrains of the density dependence of symmetry energy at the collective flow characteristic density 1.2$\rho_0$ and the $\pi^-/\pi^+$ characteristic density 1.5$\rho_0$. Panel(b): The constraints on $S_0$ and $L$ by using  elliptic flow difference, elliptic flow ratio and $\pi^-/\pi^+$ in this work (green symbols).}
\setlength{\belowcaptionskip}{0pt}
\label{Ssym-constrains}
\end{figure*}

Now, let's illustrate the constraints of symmetry energy at flow and pion characteristic densities, i.e., at 1.2$\rho_0$ and 1.5$\rho_0$. 
%Thus, by using the collective flow and pion observables\cite{YYLiu2021PRC}, one can give the constraints of suprasaturation density at two points, i.e., 1.2 $\rho_0$ and 1.5 $\rho_0$. 
In Fig.\ref{Ssym-constrains}(a), we present the values of the symmetry energy at flow characteristic density 1.2$\rho_0$ and pion characteristic density 1.5$\rho_0$. The black square symbols are the results obtained in this work, and the values of them are $S(1.2\rho_0)=34\pm 4$ MeV and $S(1.5\rho_0)=36\pm 8$ MeV. %The symbols have the same meaning as in panel (a).
The constraints of $S(\rho)$ at flow characteristic density, i.e., at 1.2$\rho_0$, are consistent with the analysis of elliptic flow ratios or elliptic flow difference by UrQMD\cite{WangYJ2014PRC} or T\"{u}QMD calculations\cite{Cozma2013PRC,Cozma2018EPJ} (which are presented by blue symbols) within statistical uncertainties. The constraints of $S(\rho)$ at pion characteristic density, i.e., at 1.5$\rho_0$, is also consistent with our previous analysis and constraints from S$\pi$RIT\cite{GJhang20}, T\"uQMD\cite{Cozma2016PLB}, dcQMD\cite{SpiRIT2021PRL} and IBUU\cite{ZGXiao2009PRL,GCYong2021PRC}  within statistical uncertainties, except the constraints obtained by LQMD\cite{ZQFeng2010PLB}. Our constraint is also consistent with the recent inclination analyses by Lynch and Betty~\cite{Lynch2022PLB} (pink shaded region) and ab initial theoretical predictions by chiral effective field theory ($\chi$EFT)~\cite{Drischler2020PRL} (orange region). %Two shaded regions are the results obtained by $\chi$EFT theory~\cite{Drischler2020PRL} and inclination analyses\cite{Lynch2022PLB}

%The difference among those calculations may be caused by using a different MDI, incompressibility $K_0$, the treatment on the threshold effects\cite{Song15} and energy conservation\cite{Cozma2016PLB} in the process of $NN\to N\Delta$. This issue should be well understood in the next step for investigating the pion and flow at subthreshold energy. % For convenience, we also present the constraints on symmetry energy at $2\rho_0$ by symbols and the density dependence of symmetry energy extracted in this work are presented within the green lines.

%obtained in this work is $37\pm 8$ MeV, and this result is also in the uncertainty of the previous constraints of symmetry energy obtained by the comparison of $\pi^-/\pi^+$ data to the IBUU \cite{ZGXiao2009PRL}, T\"{u}QMD\cite{Cozma2016PLB}, UrQMD\cite{YYLiu2021PRC}, impIBUU\cite{GCYong2021PRC} and dcQMD\cite{SpiRIT2021PRL} calculations.  

% The green lines are the boundary of density dependence of symmetry energy obtained with $L=5$ and $L=80$ MeV in this work.
%While, in the T\"{u}QMD calculations\cite{Cozma2016PLB}, the energy conservation in the process of $NN\to N\Delta$ is considered, and which modify the reaction dynamics and symmetry energy effect.

%\begin{figure}[htbp]
%\centering
%\includegraphics[angle=0,scale=0.4]{S0-L.pdf}
%\setlength{\abovecaptionskip}{0pt}
%\vspace{2em}
%\caption{The constraints on the $S_0$ and L obtained with both $v_2^n/v_2^{ch}$ and $\pi^-/\pi^+$. }
%\setlength{\belowcaptionskip}{0pt}
%\label{S0-L}
%\end{figure}

Another interesting and important question is what are the extrapolated values of symmetry energy at normal density, i.e., $S_0$ and $L$? Fig.~\ref{Ssym-constrains} (b) shows the extrapolated values of $S_0$ and $L$ at $\rho_0$ by using both $v_2^n/v_2^{ch}$ and $\pi^-/\pi^+$ in this work (black symbols). The extrapolated $S_0$ and $L$ are in $30-34$ MeV and $5-70$ MeV, respectively. The blue symbols are the results from the elliptic flow ratios, such as $v_2^n/v_2^p$, $v_2^n/v_2^H$ and $v_2^n/v_2^{ch}$\cite{Russotto2011PLB,Cozma2013PRC,WangYJ2014PRC,Russotto2016PRC,Cozma2018EPJ}, or elliptic flow difference, such as $v_2^n-v_2^p$ and $v_2^n-v_2^H$\cite{Cozma2013PRC,WangYJ2014PRC}, and dark green symbols are the results from $\pi^-/\pi^+$ ratios\cite{ZGXiao2009PRL,ZQFeng2010PLB,WJXie2013PLB,Cozma2016PLB,YYLiu2021PRC,GCYong2021PRC,SpiRIT2021PRL}. Different than the consistency at 1.2$\rho_0$ and $1.5\rho_0$ by using flow observables and pion observables, using one observable to extrapolate constraints on $S_0$ and $L$ leads to an obvious model dependence. For example, the extrapolated $L$ by UrQMD, T\"uQMD, dcQMD, IBUU04 and IBL models with only isospin sensitive pion or flow observable are different, even within the same transport model. Consequently, describing the constraints of symmetry energy at their characteristic density is  more reliable than only giving the extrapolated values of $S_0$ and $L$.

%For the extrapolated values of $S_0$ and $L$, one can expect that using the values at more than one characteristic densities will be much more reliable than only using one observable. In our work, the extrapolated values of $S_0$ and $L$ can be understood as an extraploation from 1.2$\rho_0$ and $1.5\rho_0$ since we simultaneously describe two isospin sensitive observables. 

Furthermore, we also compare the extrapolated $S_0$ and $L$ with the recent constraints by analyzing neutron skin~\cite{Reed21PRL,Pineda2021PRL,Essick21,ZhenZhang2022ARXIV}, isospin degree of freedom~\cite{Yanzhang2017PRC}, combined analysis of neutron skin, isospin diffusion and neutron stars~\cite{YXZhang2020PRC}, and combined analysis from neutron skin and astrophysics~\cite{Essick2021PRC}. Our result is below the constraints with a specific class of relativistic energy density functional\cite{Reed2021PRL}, but it is consistent with the one from the combining astrophysical data with PREX-II and chiral effective field theory, $L=53^{+14}_{-15}$ MeV\cite{Essick21}, and it is also well consistent with the constraints extracted from the charge radius of $^{54}$Ni, where they deduced $21<L<88$ MeV\cite{Pineda2021PRL}, and very recent Bayesian analysis of charge-weak form factor difference $\Delta F_{CW}$ in $^{48}$Ca and $^{208}$Pb by the CREX and PREX-2 collaborations, where they infer $-26<L<62$ Mev\cite{ZhenZhang2022ARXIV}. The consistency with the analysis of the isospin degree of freedom (IDOF) and isospin diffusion with ImQMD model\cite{Yanzhang2017PRC,YXZhang2020PRC}, and in the ab initial calculations with chiral effective field theory ($\chi$EFT)~\cite{Drischler2020PRL} and relativistic Brueckner-Hartree-Fock theory (RBHF)~\cite{SiboWang2022} is also observed.
% and from the combined analysis for  isospin diffusion, neutron skin and properties of neutron stars, $50<L<70$ MeV\cite{YXZhang2020PRC}.} %It also overlap with the very recent constraint of symmetry energy $42 < L < 117$ by using the S$\pi$RIT pion data\cite{SpiRIT2021PRL} for Sn+Sn at 270A MeV.

Even the flow and pion observables can not directly give the constraints of symmetry energy at subsaturation density, we also plot the extrapolated symmetry energy at subsaturation density with the dashed lines for understanding its validity at subsaturation density. The uncertainty of extrapolated form of symmetry energy is large, but it is still in the reasonable region and covers constraints of symmetry energy from the HIC(n/p)\cite{Morfouace2019PLB}, isospin diffusion in HIC(isodiff)\cite{Tsang2008PRL}, mass calculated by the effective Skyrme interaction\cite{Brown2013PRL} and DFT theory\cite{Kortelainen2011PRC}, Isospin analog state (IAS)\cite{Danielewicz2016NPA}, electric dipole polarization ($\alpha_D$)\cite{Zhang2015PRC} at their sensitive density, which are decoded in Ref.\cite{Lynch2022PLB}.
%and combined analysis from isospin diffusion, neutron skin and properties of neutron stars\cite{YXZhang2020PRC}. ,\textcolor {red}{from the the isospin degree of freedom (IDOF),  L = 33-61 MeV\cite{YanZhang2017PRC}, from the mass-radius relation, $L=65.2$ MeV\cite{SiboWang2022},

%Our extrapolated constraint on $L$ is smaller than the results obtained by UrQMD, TuQMD, dcQMD, IBUU04 model. 

%The dark green symbols are the results from $\pi^-/\pi^+$ ratios\cite{ZGXiao2009PRL,ZQFeng2010PLB,WJXie2013PLB,Cozma2016PLB,YYLiu2021PRC,GCYong2021PRC,SpiRIT2021PRL}. The constraint of $L$ obtained in this work is consistent with our previous constraints by using the $\pi^-/\pi^+$ ratios and properties of neutron star. 

%It is below the analysis of 

%\section{Summary and outlook}
%\label{summary}
In summary, we have investigated the influence of symmetry energy on nucleonic and pion observable, such as $v_1^{n}$, $v_1^{ch}$, $v_2^{n}$, $v_2^{ch}$, $v_2^{n}/v_2^{ch}$, $M(\pi)$ and $\pi^-/\pi^+$, with UrQMD model for Au+Au at the beam energy of 0.4A GeV. To constrain the symmetry energy at suprasaturation density with the flow and pion observables, the characteristic densities obtained with flow and pion observables are discussed. Our analysis found that the flow characteristic density is around 1.2$\rho_0$ and pion characteristic density is around 1.5$\rho_0$. By simultaneously describing the data of $v_2^n/v_2^{ch}$ and $\pi^-/\pi^+$, we got the $S(1.2\rho_0)=34\pm 4$ MeV and $S(1.2\rho_0)=36\pm 8$ MeV. The constrained symmetry energy at characteristic densities are consistent with previous analysis by using pion and flow observables with different transport models, and suggest that the reliable constraints on symmetry energy should be presented at the characteristic density of isospin sensitive observables. Further, the obtained parameter sets are also tested by simulating the pion production for Sn+Sn at 0.27A GeV. Our validated calculations show that the favored parameter sets can describe the charged pion multiplicity for three Sn+Sn reaction systems. The underestimation of $\pi^-/\pi^+$ for $^{132}$Sn+$^{124}$Sn is observed in our calculations, which may be attributed to the isospin dependent threshold effects and isospin dependent in-medium $NN\to N\Delta$ cross sections. The line of this work is in progress. 

The inconsistent results on the value of $L$ do not exactly reflect the debates, since the value of $L$ at normal density is usually extrapolated from the symmetry energy at characteristic density. To enhance the reliability of extrapolation of $S_0$ and $L$, one can expect to use more than one isospin sensitive observables which will have the information of symmetry energy at different characteristic densities. The extrapolated values of $L$ in this work are in $5-70$ MeV within $2\sigma$ uncertainty for $S_0=30-34$ MeV, which is below the analysis of PREX-II results with a specific class of relativistic energy density functional, but is consistent with the constraints from charged radius of $^{54}$Ni, from the combining astrophysical
data with PREX-II and chiral effective field theory.

\section*{Acknowledgements}
The authors thank the discussions on the transport model and symmetry energy constraints at TMEP weekly meeting. This work was supported by the National Key R\&D Program of China under Grant No. 2018 YFA0404404, the National Natural Science Foundation of China Nos.11875323, 12275359, 12205377, 11875125, U2032145, 11790320, 11790323, 11790325, and 11961141003, the Continuous Basic Scientific Research Project (No. WDJC-2019-13), the Continuous Basic Scientific Research Project (No. WDJC-2019-13), and the funding of China Institute of Atomic Energy, and the Leading Innovation Project of the CNNC under Grant No. LC192209000701, No.  LC202309000201. We acknowledge support by the computing server C3S2 in Huzhou University.

%\begin{thebibliography}{99}
%%% review paper of symmetry energy
%\bibliography{ref}

\bibliography{ref1127}

%\end{thebibliography}

\end{document}